\RequirePackage{fix-cm}

\documentclass[twocolumn,epjc3]{svjour3}  

\smartqed 
\usepackage{mathtools}
\usepackage{graphicx}
\usepackage{epstopdf}
\usepackage[version=3] {mhchem}
\usepackage{amssymb}
\usepackage{braket}
\usepackage{latexsym}
\usepackage{subfigure}
\usepackage{young}
\usepackage[dvipsnames]{xcolor}
\usepackage[toc,page]{appendix}
\usepackage{cite}
\usepackage{bbm}
\usepackage{dsfont}
\usepackage{lscape}
\usepackage{float}
\usepackage{tikz}
\usepackage{pgf}
\usetikzlibrary{matrix}
\usetikzlibrary{positioning,shapes,fit,arrows,calc}

\definecolor{myblue}{RGB}{56,94,141}

\journalname{Eur. Phys. J. A}

\begin{document}

\title{Why nuclear forces favor the highest weight irreducible representations of the fermionic SU(3) symmetry
}

\author{Andriana Martinou\thanksref{e1,addr1}
        \and
        Dennis Bonatsos\thanksref{addr1}
\and
K. E. Karakatsanis\thanksref{addr1,addr2}
\and
S. Sarantopoulou\thanksref{addr1}
\and
I.E. Assimakis\thanksref{addr1}
\and
S.K. Peroulis\thanksref{addr1}
\and
N. Minkov\thanksref{addr3}
}

\thankstext{t1}{This research is co-financed by Greece and the European Union (European Social Fund- ESF) through the Operational Programme ``Human Resources Development, Education and Lifelong Learning 2014-2020" in the context of the project ``Nucleon Separation Energies" (MIS 5047793). NM acknowledges support by the Bulgarian National Science Fund (BNSF) under Contract No.KP-06-N48/1}
\thankstext{e1}{e-mail: martinou@inp.demokritos.gr}

\authorrunning{A. Martinou et al.} 

\institute{Institute of Nuclear and Particle Physics, National Centre of Scientific Research ``Demokritos'', GR-15310 Aghia Paraskevi, Attiki, Greece. \label{addr1}
           \and
Department of Physics, Faculty of Science, University of Zagreb, HR-10000 Zagreb, Croatia.\label{addr2}
\and
          Institute for Nuclear Research and Nuclear Energy, Bulgarian Academy of Sciences, 72 Tzarigrad Road, 1784 Sofia, Bulgaria.  \label{addr3} }

\date{Received: date / Accepted: date}

\maketitle

\begin{abstract}
The consequences of the attractive, short--range nucleon--nucleon (NN) interaction on the wave functions of the Elliott SU(3) and the proxy-SU(3) symmetry are discussed. The NN interaction favors the most symmetric spatial SU(3) irreducible representation, which corresponds to the maximal spatial overlap among the fermions. The percentage of the symmetric components out of the total in an SU(3) wave function is introduced, through which it is found, that no SU(3) irrep is more symmetric than the highest weight irrep for a certain number of valence particles in a three dimensional, isotropic, harmonic oscillator shell. The consideration of the highest weight irreps in nuclei and in alkali metal clusters, leads to the prediction of a prolate to oblate shape transition beyond the mid--shell region.
\end{abstract}

\section{Introduction}\label{intro}

The recent introduction of the proxy-SU(3) Model \cite{proxy1,proxy2,proxy4} triggered a stormy question: Why is the highest weight irreducible representation (irrep) of SU(3) used instead of the irrep with the highest value of the second order Casimir operator of SU(3), which corresponds to the maximum value of the quadrupole--quadrupole interaction?  It is the purpose of the present work to answer this question, exposing all the physical concepts and mathematical techniques needed. 

Symmetries play an important role in the description of physical systems, especially in cases in which they can provide parameter--independent predictions of general validity. In parallel, the harmonic oscillator is occupying a central place in many branches of physics, being in many cases a very good approximation to the potential describing the physical system. Finite shells appearing in the case of the three--dimensional harmonic oscillator (3D-HO) are known to possess $U(\Omega)$ symmetries with $\Omega=3, 6, 10, 15, 21, 28, \dots$  having SU(3) subalgebras \cite{Wybourne,Moshinsky,IacLie}. 

The SU(3) symmetry is playing a central role in the description of nuclear shapes and spectra \cite{Kota}, since its introduction by Elliott \cite{Elliott1,Elliott2,Elliott3,Elliott4} for the exact description of light nuclei, extended later to heavy nuclei by various approximations, including the pseudo-SU(3) Model \cite{pseudo1,pseudo2,DW1,Quesne}, the quasi-SU(3) Model \cite{Zuker1,Zuker2}, and, more recently, the proxy-SU(3) Model \cite{proxy1,proxy2,proxy3,proxy4}. In addition, several algebraic models containing SU(3) as one of their limiting symmetries have been introduced, including the Interacting Boson Model (IBM) \cite{IA,IVI,FVI} and the Vector Boson Model (VBM) \cite{Raychev,Afanasev,RR27}, which use bosons as their building blocks, as well as the Fermion Dynamic Symmetry Model (FDSM) \cite{FDSM} and the Symplectic Model \cite{Rosensteel1979,Rowe1980,RW}, which use fermions. Similar in spirit algebraic models bearing SU(3) limiting symmetries, like the Vibron Model \cite{IL,FVI},  have also been introduced for the description of diatomic and polyatomic molecules. 

Deformed nuclei have also been described in the framework of the Nilsson Model \cite{Nilsson1,Nilsson2}, which consists of a 3D-HO with cylindrical symmetry to which the spin--orbit interaction \cite{Mayer1} is added, which is known to be essential for the reproduction of the experimentally seen nuclear magic numbers 2, 8, 20, 28, 50, 82, 126, \dots \cite{Mayer}. The same model, without the spin--orbit interaction, has been found very successful \cite{Clemenger} for the description of atomic clusters \cite{deHeer,Brack,Nester,deHeer2} and for reproducing the experimentally seen magic numbers, which in the special case of alkali metal clusters are 2, 8, 20, 40, 58, 92, 138, 198, \dots  \cite{Martin1,Martin2,Bjorn1,Bjorn2,Knight1,Peder,Brec1,Brec2}.

Similarities observed in atomic nuclei and atomic clusters have been investigated \cite{Nester,Greiner} since the experimental identification of the latter \cite{deHeer,deHeer2}. Transitions from prolate (rugby--ball like) to oblate (pancake--like) deformed shapes have been observed both in atomic nuclei\cite{Namenson,Alkhomashi,Wheldon,Podolyak,Linnemann}
and in alkali metal clusters \cite{Borggreen,Pedersen1,Pedersen2,Haberland,Schmidt}. We are going to show, that these transitions can be explained by the dominance of the highest weight (hw) spatial irreducible representations (irreps) \cite{code,Wybourne} of SU(3), which is due to the attractive, short--range nature of the nucleon--nucleon interaction. We are also going to clarify the physical content of the spatial hw irreps by showing, that these irreps are bearing the maximum amount of spatial symmetrization for a given number of fermions. A by--procuct of the dominance of the spatial hw irreps of SU(3) is the dominance of prolate over oblate deformed shapes in the ground states of even--even nuclei, which has been a long--standing puzzle \cite{Hamamoto} in Nuclear Physics. It will become obvious, that all these effects are rooted in the dominance of the spatial highest weight irreps in finite fermionic shells, the relevant predictions being completely independent of any free parameters.

\section{Basic properties of the strong force}\label{force}

The strong force is applied among nucleons and binds them together into the nucleus. This force derives from fundamental interactions among quarks and gluons obeying to the equations of Quantum Chromodynamics \cite{Chromo}. Unfortunately these equations have not been solved and so the NN interaction remains unknown. 

Yet general properties of effective potentials, which resemble the NN interaction are known. High precision NN potentials are available \cite{Machleidt2001,Wiringa1995,Stoks1994}, which respect some general characteristics of the NN interaction at different length scales \cite{Aoki2010}:\\
1. At relevant nucleon--nucleon distances $d>2\mbox{ fm}$ the tensor force dominates, which has a spin--isospin dependence \cite{Arima1960,Schmidt2020}.\\
2. At the short range of $1\mbox{ fm} < d < 2 \mbox{ fm}$ a spin--isospin independent attraction binds the nucleons together \cite{Jastrow1951,Schmidt2020}.\\
3. At the extremely short distances of $d< 1\mbox{ fm}$ a strong repulsive core appears mainly due to the Pauli principle \cite{Pauli} even among protons and neutrons due to their constituents \cite{Faessler1988,Fernandez2020,Jastrow1951}.

These general properties of the strong force have been taken into account in the Wigner SU(4) symmetry \cite{Wigner1937} and in the Elliott SU(3) symmetry \cite{Elliott1,Elliott2}. Specifically the tensor force is included in the Shell Model \cite{Mayer} and its extension, which is the Elliott SU(3) symmetry, through the spin--orbit interaction \cite{Arima1960,Elliott4}. The spin--isospin independent attraction is well treated in the Wigner SU(4) symmetry and as we will show in this article, it is decisive for the favored Elliott SU(3) irrep. Finally the repulsive core at extremely short inter-nucleon distances is included in those symmetry models through the Pauli Exclusion Principle \cite{Pauli}.

\section{The many--body wave functions}\label{wf}

The Shell Model \cite{Mayer1} is widely accepted, to describe in the microscopic level the atomic nuclei. A basic assumption of the model is, that the nucleons are subjected to a mean field potential, which may be represented by the three dimensional isotropic harmonic oscillator (3D-HO) plus the spin--orbit interaction, leading to the single--particle Hamiltonian for the $i^{th}$ nucleon:
\begin{equation}\label{H}
h_i={\mathbf{p_i}^2\over 2M}+{1\over 2}M\omega^2 \mathbf{r_i}^2+\upsilon_{l_is_i}\hbar\omega \mathbf{l}_i\cdot \mathbf{s}_i,
\end{equation}
where the first two terms of Eq. (\ref{H}) represent the three dimensional isotropic harmonic oscillator:
\begin{equation}\label{h0}
h_{0,i}={\mathbf{p_i}^2\over 2M}+{1\over 2}M\omega^2 \mathbf{r_i}^2
\end{equation}
with $\mathbf{p_i}$, $\mathbf{r_i}$, $M$, $\omega$ being the momentum, spatial coordinate, mass and oscillation frequency respectively, while the last term is the spin--orbit interaction, with $\mathbf{l_i}$, $\mathbf{s_i}$ being the orbital angular momentum and spin, and $\upsilon_{l_is_i}$ is a strength parameter \cite{Nilsson2} (see Table I of Ref. \cite{proxy1} for the values). An $l_i^2$ term is usually added in the above Hamiltonian, which serves for the flattening of the mean field potential. 

The greatest success of the Shell Model has been the prediction of the nuclear magic numbers 2, 8, 20, 28, 50, 82, 126. Despite this major success the Shell Model has been confronted with skepticism in the early years of its introduction. The main problem was, that the short--range character of the NN interaction means, that one cannot use a smooth mean field potential in the single--particle Hamiltonian \cite{Elliott1957}. This obstacle has been overpassed theoretically through the Pauli Exclusion Principle \cite{Pauli,Fermi}, to which the nucleons, being fermions, obey. This principle dictates, that only one fermion at a time may occupy a given state. Thus despite of the feeling of a fluctuating potential, each nucleon has a smooth path inside the nucleus.

Upon the theoretical approval of the Shell Model J. P. Elliott has proved, that a nuclear shell consisting of single--particle orbitals with common number of oscillator quanta $\mathcal{N}$ possesses an SU(3) symmetry \cite{Elliott1,Elliott2}. The building blocks of the Elliott SU(3) Model are the eigenstates of the Hamiltonian of Eq. (\ref{H}), {\it i.e.}, the Shell Model orbitals. 

If the spherical coordinate system is used, the eigenstates are labeled as $\ket{n,l,j,m_j}$, where $n$ is the radial quantum number getting values $n=0,1,2,...$ and obeying the equation \cite{Davies}:
\begin{equation}
\mathcal{N}=2n+l,
\end{equation}
$j$ is the total angular momentum, which derives after the spin--orbit coupling $\bf{j}=\bf{l}+\bf{s}$ and $m_j$ is the projection of $j$, having integer values within the interval $-j\le m_j\le j$ \cite{Mayer1}. 

If the cartesian coordinate system is to be used, then the eigenstates of the $h_{0,i}$ of Eq. (\ref{h0})
are labeled as $\ket{n_z,n_x,n_y,m_s}$, with $n_z,n_x,n_y$ being the number of quanta in each cartesian direction $z,x,y$ respectively and $m_s=\pm {1\over 2}$ is the projection of the spin $s={1\over 2}$ of the nucleon \cite{proxy4}. A unitary transformation can be applied among the spherical and the cartesian states $\ket{n_\rho,l,j,m_j}\leftrightarrow \ket{n_z,n_x,n_y,m_s}$ \cite{proxy4}. The cartesian states $\ket{n_z,n_x,n_y,m_s}$ are convenient for the calculation of the Elliott SU(3) irreps $(\lambda,\mu)$ \cite{Elliott1,Elliott2,code} and thus they have been chosen as the intrinsic states of the Elliott SU(3) Model \cite{Elliott3,Harvey,Wilsdon}.

The many--particle wave function is simply a Slater determinant \cite{Slater,Ring} of the $\ket{n_z,n_x,n_y,m_s}$ states \cite{Wilsdon}, which represents a totally antisymmetric many--particle wave function, as dictated by the Pauli principle for a fermion system. This complicated wave function combines both the spatial and the spin--isospin information for all the valence nucleons. Fortunately the overall, antisymmetric wave function can be decomposed into a spatial and a spin--isospin part for the many nucleon system.

The spatial part refers to 3D-HO shell, which consists of orbitals of $\mathcal{N}$ number of quanta, possesses $\Omega= {(\mathcal{N}+1)(\mathcal{N}+2)\over 2}$ spatial orbitals, which in the cartesian coordinates are written as:
\begin{gather}\label{order}
\ket{n_z,n_x,n_y}:\ket{\mathcal{N},0,0}, \ket{\mathcal{N}-1,1,0}, \ket{\mathcal{N}-1,0,1},\nonumber\\ \ket{\mathcal{N}-2,2,0},\ket{\mathcal{N}-2,1,1},\ket{\mathcal{N}-2,0,2},...,\ket{0,0,\mathcal{N}}.\nonumber\\
\end{gather}
The symmetry of this set of spatial orbitals is $U(\Omega)$ \cite{Elliott1,Elliott2}. Young patterns of $U(\Omega)$ symmetry for proton and neutron configurations have boxes, which represent the particles, arranged in $\Omega$ rows and 4 columns (for proton and neutron configurations) and they are described by the partition:
\begin{equation}\label{Y1}
[f_1,f_2,...,f_{\Omega}],
\end{equation}
with $f_1\ge f_2\ge ...\ge f_{\Omega}$. The numbers $f_1,f_2,...,f_{\Omega}$ are the number of boxes in each row.

Each of the orbitals can be occupied by 2 protons and 2 neutrons with opposite spin projections. The isospin of a nucleon is $t={1\over 2}$ and its projection is  $m_t={1\over 2}$, if it is a proton, and $m_{t}=-{1\over 2}$, if it is a neutron. The spin--isospin many--particle wave function has a $U(4)$ symmetry, usually called Wigner's SU(4) symmetry \cite{Wigner1937}. For the short--range territory of a spin--isospin independent attraction Wigner at Ref. \cite{Wigner1937} and Hund at \cite{Hund1937} used a Hamiltonian with SU(4) symmetry, which did not involve the ordinary spin and applied equal forces among all nucleons (protons and neutrons). The irreps of the $U(4)$ symmetry are \cite{DraayerBook,Kota2018}:
\begin{equation}\label{con}
[f_1^c,f_2^c,f_3^c,f_4^c],
\end{equation}
where $c$ stands for the conjugate irreps of the $U(\Omega)$ symmetry, since the pattern (\ref{con}) has as rows the columns of the pattern (\ref{Y1}).  Obviously the Young pattern, which corresponds to (\ref{con}), has boxes in four rows. Each box represents a valence nucleon. For a nucleus with neutron excess $N_{val}\ge Z_{val}$ the $f_1^c$ boxes in the first row are neutrons with $m_t=-{1\over 2},m_s=+{1\over 2}$, the $f_2^c$ boxes in the second row represent neutrons with $m_t=-{1\over 2},m_s=-{1\over 2}$, the $f_3^c$ boxes in the third row are protons with $m_t=+{1\over 2},m_s=+{1\over 2}$, while the $f_4^c$ boxes in the forth row have $m_t=+{1\over 2},m_s=-{1\over 2}$.

The combination of the spatial symmetry with that of the spin--isospin is labeled as \cite{Talmi}:
\begin{equation}
U(\Omega)\otimes U(4)=U(4\Omega).
\end{equation}
The combined wave functions with $U(4\Omega)$ symmetry, which include the spatial, spin and isospin information for the many nucleon problem, are the Slatter determinants \cite{Wilsdon} and according to the Pauli principle \cite{Pauli} they are totally antisymmetric.

\section{The binding energy in the SU(4) symmetry}

The Majorana operator of the exchange of two particles $1\leftrightarrow 2$ in the spatial coordinates of a two--body wave function $\phi({\vec r_{1}},{\vec r_2})$ is \cite{Talmi}:
\begin{equation}
\hat P_{12}^x\phi({\vec r_{1}},{\vec r_2})=\phi({\vec r_{2}},{\vec r_1}).
\end{equation}
The eigenvalue of the above operator for a spatially symmetric wave function is $+1$, while for an antisymmetric one is $-1$.
For a many--body wave function, the eigenvalue of the Majorana operator $\hat P^x$ is \cite{Talmi}:
\begin{equation}\label{P}
P^x=\sum_{i<i'}P_{ii'}^x=P_S-P_A,
\end{equation}
where $(i,i')$ are the nucleon pairs and the term
\begin{equation}
P_S=f_2^c+2f_3^c+3f_4^c
\end{equation}
counts the number of the symmetric pairs of particles in the spatial wave function with $U(\Omega)$ symmetry, while the 
\begin{eqnarray}
P_A={1\over 2}(f_1^c(f_1^c-1)+f_2^c(f_2^c-1)\nonumber\\
+f_3^c(f_3^c-1)+f_4^c(f_4^c-1))
\end{eqnarray}
is the number of antisymmetric pairs in the spatial coordinates.

Using a Majorana two--body force \cite{Talmi}:
\begin{equation}
P_{12}^xV_M(|r_1-r_2|)
\end{equation}
in the Hamiltonian, the ground state binding energy due to this force in the many nucleon problem arises to be \cite{Talmi,Franzini1963,Isacker1997}:
\begin{equation}\label{E}
BE=a(A)-b(A)P^x,
\end{equation}
 $a(A), b(A)$ are parameters, which depend on the mass number $A$, with $a(A)$ being positive and $b(A)$ being negative.

It holds therefore, that the more symmetric the spatial wave function is, the greater is the value of Eq. (\ref{P}) and the greater is the binding energy of Eq. (\ref{E}). A test of the validity of the Wigner SU(4) symmetry has been performed in Refs. \cite{PVIWigner,Nayak2001}, while more details can be found in Ref. \cite{Isacker1997}. From the Wigner SU(4) symmetry arises, that the favored irrep is the most spatially symmetric. In other words, the more spatially symmetric is the nuclear state, the more bound is the nucleus.

\section{The Elliott SU(3) irreps}\label{sec1}

The Elliott SU(3) irreps $(\lambda,\mu)$ can be determined, by the distribution of the particles in the valence orbitals. For instance the $pf$ shell with $\mathcal{N}=3$ number of quanta contains 10 spatial orbitals of the type $\ket{n_z,n_x,n_y}$:
\begin{eqnarray}\label{order3}
\ket{3,0,0},\ket{2,1,0},\ket{2,0,1},\ket{1,2,0},\ket{1,1,1},\nonumber\\
\ket{1,0,2},\ket{0,3,0},\ket{0,2,1},\ket{0,1,2},\ket{0,0,3}.
\end{eqnarray}
This shell possesses a $U(10)$ symmetry, where ``10" refers to the number of orbitals and has a capacity of 20 protons or neutrons. Supposing for example, that the $pf$ shell contains 10 protons and that each orbital is occupied by 2 particles, the corresponding spatial Young diagram of the $U(10)$ symmetry is:
\begin{equation}\label{10}
\begin{Young} &\cr&\cr&\cr&\cr&\cr \end{Young},
\end{equation}
where each box represents a proton. In general the spatial Young diagram of a 3D-HO shell with $\mathcal{N}$ quanta and $U(\Omega)$ symmetry, has at most 2 columns for configurations of identical nucleons, since at most two of them with opposite spin projections may occupy a certain orbital. The irrep of the spatial $U(\Omega)$ symmetry is labeled by (\ref{Y1}) and for the Young diagram of Eq. (\ref{10}) it is $[2,2,2,2,2,0,0,0,0,0]$ or $[2^5]$.

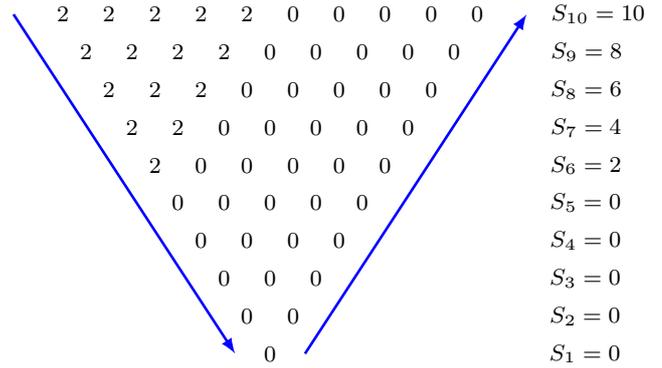
\begin{figure}
\begin{tikzpicture}[line width=1pt,>=latex]
\node (a0) {};
\node[right=0.2cm of a0] (a1) {2};
\node[right=0.2cm of a1] (a2) {2};
\node[right=0.2cm of a2] (a3) {2};
\node[right=0.2cm of a3] (a4) {2};
\node[right=0.2cm of a4] (a5) {2};
\node[right=0.2cm of a5] (a6) {0};
\node[right=0.2cm of a6] (a7) {0};
\node[right=0.2cm of a7] (a8) {0};
\node[right=0.2cm of a8] (a9) {0};
\node[right=0.2cm of a9] (a10) {0};
\node[right=0.2cm of a10] (a11) {};
\node[right=0.2cm of a11] (a12) {$S_{10}=10$};

\node at ($(a1)!0.5!(a2)+(0,-0.5)$) (b1) {2};
\node at ($(a2)!0.5!(a3)+(0,-0.5)$) (b2) {2};
\node at ($(a3)!0.5!(a4)+(0,-0.5)$) (b3) {2};
\node at ($(a4)!0.5!(a5)+(0,-0.5)$) (b4) {2};
\node at ($(a5)!0.5!(a6)+(0,-0.5)$) (b5) {0};
\node at ($(a6)!0.5!(a7)+(0,-0.5)$) (b6) {0};
\node at ($(a7)!0.5!(a8)+(0,-0.5)$) (b7) {0};
\node at ($(a8)!0.5!(a9)+(0,-0.5)$) (b8) {0};
\node at ($(a9)!0.5!(a10)+(0,-0.5)$) (b9) {0};
\node[right =0.2cm of b9](b10){};
\node[right=0.5cm of b10] (b11) {$S_{9}=8$};

\node at ($(b1)!0.5!(b2)+(0,-0.5)$) (c1) {2};
\node at ($(b2)!0.5!(b3)+(0,-0.5)$) (c2) {2};
\node at ($(b3)!0.5!(b4)+(0,-0.5)$) (c3) {2};
\node at ($(b4)!0.5!(b5)+(0,-0.5)$) (c4) {0};
\node at ($(b5)!0.5!(b6)+(0,-0.5)$) (c5) {0};
\node at ($(b6)!0.5!(b7)+(0,-0.5)$) (c6) {0};
\node at ($(b7)!0.5!(b8)+(0,-0.5)$) (c7) {0};
\node at ($(b8)!0.5!(b9)+(0,-0.5)$) (c8) {0};
\node[right =0.2cm of c8](c9){};
\node[right=0.8cm of c9] (c10) {$S_{8}=6$};

\node at ($(c1)!0.5!(c2)+(0,-0.5)$) (d1) {2};
\node at ($(c2)!0.5!(c3)+(0,-0.5)$) (d2) {2};
\node at ($(c3)!0.5!(c4)+(0,-0.5)$) (d3) {0};
\node at ($(c4)!0.5!(c5)+(0,-0.5)$) (d4) {0};
\node at ($(c5)!0.5!(c6)+(0,-0.5)$) (d5) {0};
\node at ($(c6)!0.5!(c7)+(0,-0.5)$) (d6) {0};
\node at ($(c7)!0.5!(c8)+(0,-0.5)$) (d7) {0};
\node[right =0.2cm of d7](d8){};
\node[right=1.1cm of d8] (d9) {$S_{7}=4$};

\node at ($(d1)!0.5!(d2)+(0,-0.5)$) (e1) {2};
\node at ($(d2)!0.5!(d3)+(0,-0.5)$) (e2) {0};
\node at ($(d3)!0.5!(d4)+(0,-0.5)$) (e3) {0};
\node at ($(d4)!0.5!(d5)+(0,-0.5)$) (e4) {0};
\node at ($(d5)!0.5!(d6)+(0,-0.5)$) (e5) {0};
\node at ($(d6)!0.5!(d7)+(0,-0.5)$) (e6) {0};
\node[right =0.2cm of e6](e7){};
\node[right=1.4cm of e7] (e8) {$S_{6}=2$};

\node at ($(e1)!0.5!(e2)+(0,-0.5)$) (f1) {0};
\node at ($(e2)!0.5!(e3)+(0,-0.5)$) (f2) {0};
\node at ($(e3)!0.5!(e4)+(0,-0.5)$) (f3) {0};
\node at ($(e4)!0.5!(e5)+(0,-0.5)$) (f4) {0};
\node at ($(e5)!0.5!(e6)+(0,-0.5)$) (f5) {0};
\node[right =0.2cm of f5](f6){};
\node[right=1.7cm of f6] (f7) {$S_{5}=0$};

\node at ($(f1)!0.5!(f2)+(0,-0.5)$) (g1) {0};
\node at ($(f2)!0.5!(f3)+(0,-0.5)$) (g2) {0};
\node at ($(f3)!0.5!(f4)+(0,-0.5)$) (g3) {0};
\node at ($(f4)!0.5!(f5)+(0,-0.5)$) (g4) {0};
\node[right =0.2cm of g4](g5){};
\node[right=2.0cm of g5] (g6) {$S_{4}=0$};

\node at ($(g1)!0.5!(g2)+(0,-0.5)$) (h1) {0};
\node at ($(g2)!0.5!(g3)+(0,-0.5)$) (h2) {0};
\node at ($(g3)!0.5!(g4)+(0,-0.5)$) (h3) {0};
\node[right =0.2cm of h3](h4){};
\node[right=2.3cm of h4] (h5) {$S_{3}=0$};

\node at ($(h1)!0.5!(h2)+(0,-0.5)$) (i1) {0};
\node at ($(h2)!0.5!(h3)+(0,-0.5)$) (i2) {0};
\node[right =0.2cm of i2](i3){};
\node[right=2.6cm of i3] (i4) {$S_{2}=0$};

\node at ($(i1)!0.5!(i2)+(0,-0.5)$) (k1) {0};
\node[right =0.2cm of k1](k2){};
\node[right=2.9cm of k2] (k3) {$S_{1}=0$};

\node[left = 0.01 of k1](k0){};
\node[right = 0.01 of k1](k2){};

\draw [->,blue] (a0.west) to (k0.west);
\draw [->,blue] (k2.east) to (a11.east);

\end{tikzpicture}
\caption{A possible Gelfand--Zeitlin pattern \cite{Gelfand,code,Martinou2018} for 10 protons/neutrons in the $pf$ shell with $U(10)$ symmetry. The numbers must not increase along the directions of the arrows and from the left to the right side of each row. The $S_{10},S_9,...,S_1$ are the summations of the numbers of each row, which lead to the calculation of the weight vector of Eq. (\ref{weight}). }\label{hw10}
\end{figure}

The distribution of the valence particles into the valence space is handled mathematically by the reduction $U(\Omega)\supset U(\Omega-1)\supset U(\Omega-2)\supset...\supset U(1)$, which is labeled by the Gelfand-Zeitlin (GZ) patterns \cite{Gelfand}. Such patterns are presented in Eqs. (1), (7), (8) of Ref. \cite{code} and in Fig. \ref{hw10} and they look like upside triangles. The upper row of the triangle is the partition (\ref{Y1}). The next row is the partition $[f_1,f_2,...,f_{\Omega-1}]$ of the reduced $U(\Omega-1)$ algebra and so on till the bottom row for the $U(1)$ algebra. One of the possible GZ patterns for 10 protons in the $pf$ shell is presented in Fig. \ref{hw10}.

There are numerous possible distributions of the particles in the valence space. Each particle distribution in the $\ket{n_z,n_x,n_y}$ states is labeled by the weight vector of the corresponding GZ pattern. If the summations of the numbers of each row in the GZ pattern are labeled $S_\Omega, S_{\Omega-1},...,S_1$ (heading from the top to the bottom of the pattern as in Fig. \ref{hw10}), then the weight vector is \cite{code}:
\begin{eqnarray}\label{weight}
w=(S_{\Omega}-S_{\Omega-1},S_{\Omega-1}-S_{\Omega-2},...,S_1).
\end{eqnarray}
The difference $S_{\Omega}-S_{\Omega-1}$ reflects to the number of particles, that have been placed in the cartesian orbital $\ket{n_z,n_x,n_y}=\ket{\mathcal{N},0,0}$ of Eq. (\ref{order}), the second difference $S_{\Omega-1}-S_{\Omega-2}$ to the number of particles in the $\ket{n_z,n_x,n_y}=\ket{\mathcal{N}-1,1,0}$ orbital and so on till $S_1$, which corresponds to the occupancy of the $\ket{n_z,n_x,n_y}=\ket{0,0,\mathcal{N}}$. The highest weight irrep is the one, which corresponds to a weight vector with the maximum values, allowed by the Pauli principle, in the first coordinates of the vector $w$ as defined in Eq. (\ref{weight}). For a proton or neutron configuration the maximum value in a coordinate of the $w$ vector is 2, while for a proton and neutron configuration the maximum value is 4. Thus the GZ pattern of Fig. \ref{hw10} has the highest weight vector 
\begin{equation}\label{w10}
w=(2,2,2,2,2,0,0,0,0,0),
\end{equation}
which means, that in this example each of the first five orbitals of Eq. (\ref{order3}) are occupied by two protons. Note, that the summation of the coordinates of the $w$ vector equals to the number of valence particles.

In the Elliott SU(3) Model the spatial $U(\Omega)$ algebra of a 3D-HO shell has a $U(3)$ subalgebra \cite{Elliott1,Elliott2}, which itself reduces to an SU(3) algebra:
\begin{equation}\label{subset}
U(\Omega)\supset U(3) \supset SU(3).
\end{equation} 
In the above the number ``3" stands for the three cartesian directions $z,x,y$. The number of boxes in each of the three rows of a Young diagram of the above U(3) symmetry reflects to the summations for every valence proton/neutron ($i$) of the quanta in each cartesian direction $z,x,y$:
\begin{equation}\label{summations}
\sum_{i}n_{z,i},\quad \sum_{i}n_{x,i},\quad \sum_in_{y,i},
\end{equation}
respectively.
Since quanta are bosons, one may place infinite number of boxes in the rows of the Elliott $U(3)$ Young diagram with partition:
\begin{equation}\label{fi}
[f_1,f_2,f_3],\mbox{ with }f_1\ge f_2\ge f_3.
\end{equation}
If $\sum_{i}n_{z,i}\ge \sum_{i}n_{x,i}\ge \sum_in_{y,i}$, then the spatial $U(3)$ partition is:
\begin{equation}\label{U(3)}
[f_1,f_2,f_3]=[\sum_{i}n_{z,i},\sum_{i}n_{x,i},\sum_in_{y,i}].
\end{equation}
Since the weight vector of Eq. (\ref{w10}) indicates that 2 particles are placed in the first 5 orbitals of (\ref{order3}), the summations of Eq. (\ref{U(3)}) are:
\begin{gather}
\sum_{i=1}^{10}n_{z,i}=2(3+2+2+1+1)=18,\\
\sum_{i=1}^{10}n_{x,i}=2(0+1+0+2+1)=8,\\
\sum_{i=1}^{10}n_{y,i}=2(0+0+1+0+1)=4
\end{gather}
and thus for the relevant example the $U(3)$ irrep is:
\begin{equation}\label{ex}
[f_1,f_2,f_3]=[18,8,4].
\end{equation}
A fully filled column in the $U(3)$ Young diagram may be erased \cite{Lipas}:
\begin{equation}
[f_1,f_2,f_3]=[f_1-f_3,f_2-f_3,0].
\end{equation}
Consequently the Young diagram of the U(3) irrep of Eq. (\ref{ex}) is:
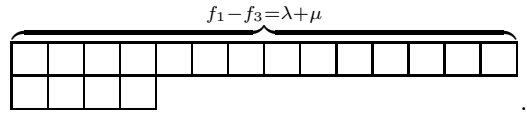
\begin{equation*}
\overbrace{\begin{Young}
&&&&&&&&&&&&&\cr
&&&\cr
\end{Young}}^{f_1-f_3=\lambda+\mu}.
\end{equation*}

The Elliott $SU(3)$ irrep $(\lambda,\mu)$ is given by \cite{Elliott1,Elliott2}:
\begin{gather}
\lambda=f_1-f_2,\label{l}\\
\mu=f_2-f_3,\label{m}
\end{gather} 
which for the example of Eq. (\ref{ex}) gives $(\lambda,\mu)=(10,4)$. The irrep $(\lambda,\mu)$ reflects to a spatial, many--quanta wave function with a total number of $\lambda+2\mu$ quanta \cite{Elliott2}. In general the symmetry of the wave function is described by the partition $[\lambda+\mu,\mu]$ \cite{Elliott2}. Such a wave function transforms as a tensor of rank $\lambda+2\mu$ \cite{Elliott2}. The $\lambda+\mu$ components out of the total are symmetric upon their interchange \cite{Elliott2}, while the $\mu$ are nor symmetric neither antisymmetric. 

For clarity we present a simpler example. If three protons are placed in the $p$ nuclear shell with $\mathcal{N}=1$ according to the highest weight vector $w=(2,1,0)$, then two of them are placed in the orbital $\ket{n_z,n_x,n_y}=$ $\ket{1,0,0}$ and one in the $\ket{0,1,0}$ cartesian orbital.
Therefore the $U(3)$ partition, which results from Eq. (\ref{U(3)}), is $[2,1,0]$.
This state may be represented by the $U(3)$ Young pattern:
\begin{equation}\label{123}
\begin{Young}&\cr\cr \end{Young},
\end{equation}
and has an SU(3) irreducible representation (irrep) $(\lambda,\mu)$ $=(1,1)$ according to Eqs. (\ref{l}), (\ref{m}). If $a^\dagger_\alpha(q)$ is the boson creation operator \cite{Cohen}, which gives to the cartesian direction $\alpha=z,x,y$ the $q^{th}$ quantum, then the terms of the spatial many--quanta wave function of this example are of the type \cite{Harvey}:
\begin{equation}
a_z^\dagger(1)a_z^\dagger(2)a_x^\dagger(3)\ket{0},
\end{equation}
with $\ket{0}$ being the vacuum state, namely the $1s$ Shell Model orbital \cite{Mayer1,Mayer}. The quantum--number Young tableaux, which represent the spatial, many--quanta wave function of this example are \cite{Lipas}:
\begin{equation}\label{Young1}
\begin{Young}
z&z\cr
x\cr\end{Young}\mbox{  ,  }
\begin{Young}1&2\cr3\cr
\end{Young}
\end{equation}
where the z, x, y represent a quantum in each cartesian direction, while the numbers $1,2,3$ enumerate the quanta and can be placed in the Young pattern so as the numbers increase from the left to the right and downwards. The wave function of the quantum--number Young tableaux of (\ref{Young1}) is \cite{Lipas}:
\begin{eqnarray}\label{ex2}
\Phi_{spatial}=\sqrt{1\over 6}(2\phi_z(1)\phi_z(2)\phi_x(3)-\phi_z(1)\phi_x(2)\phi_z(3)\nonumber\\-\phi_x(1)\phi_z(2)\phi_z(3)),\nonumber\\
\end{eqnarray}
where $\phi_\alpha(q)$ is a Hermite polynomial with the $q^{th}$ quantum in the $\alpha=z,x,y$ direction \cite{Cohen}. Obviously this wave function is symmetric upon the transposition $1\leftrightarrow 2$, but there is no symmetry in the transpositions $1\leftrightarrow 3$ and $2\leftrightarrow 3$. Indeed only two quanta are symmetric upon their interchange in Eq. (\ref{ex2}), while the third quantum is nor symmetric neither antisymmetric. The number of symmetric quanta for the above example is $\lambda+\mu=2$.

This is clearly stated in Ref. \cite{Greiner1989}, where the spatial wave function for the Young pattern of Eq. (\ref{123}), is labeled:
\begin{equation}
\Phi_{spatial}=\hat S_{q,q'}\hat A_{{q',q''}}\Phi(1,2,3),
\end{equation}
where $\hat S_{q,q'}$ is the symmetrizer operator of the $q,q'$ quanta, $\hat A_{q',q''}$ is the antisymmetrizer operator of the $q',q''$ quanta and $\Phi(1,2,3)=\phi_z(1)\phi_z(2)\phi_x(3)$ for the example of Eq. (\ref{Young1}). It is true, that if a state is antisymmetrized in $q',q''$ and thereafter is antisymmetrized in $q,q'$, then the antisymmetry of $q',q''$ is lost. Consequently the operator, which is applied last, controls the result \cite{Greiner1989}.

Finally a $(\lambda,\mu)$ irrep with $\mu>0$ corresponds to a wave function with mixed symmetry, while a $(\lambda,0)$ irrep to a totally symmetric spatial state. We define the ratio:
\begin{equation}\label{r}
r(\lambda,\mu)={\lambda+\mu\over\lambda+2\mu}\cdot 100\%,
\end{equation}
which measures the percentage of the symmetric quanta $\lambda+\mu$ out of the total number of quanta $\lambda+2\mu$.

\section{The $QQ$ interaction}

The overall $QQ$ interaction in the Elliott SU(3) Model \cite{Elliott1,Elliott2,Elliott3} is determined through:
\begin{equation}\label{QQ}
QQ=4C_2-3L(L+1),
\end{equation}
where $L^2$ is the eigenvalue of the squared angular momentum operator:
\begin{equation}
\hat L^2=\sum_i\hat l_i^2
\end{equation}
and $\hat C_2$ is the second order Casimir operator of SU(3) with the eigenvalue \cite{Elliott4}:
\begin{equation}
C_2=\lambda^2+\mu^2+\lambda\mu+3(\lambda+\mu),
\end{equation}
or
\begin{equation}\label{C2}
C_2=(\lambda+\mu)^2+3(\lambda+\mu)-\lambda\mu.
\end{equation}

The nuclear quadrupole deformation parameter $\beta$ of the Bohr--Mottelson Model \cite{BohrII} is connected with the $C_2$ as \cite{Castanos}:
\begin{equation}
\beta^2={4\pi\over 5(A\bar r^2)^2}(C_2+3),
\end{equation}
with $A$ being the mass number and $\bar r^2=0.87^2A^{1/3}$ is the dimensionless mean square radius. Thus for a certain nucleon number in a given valence 3D-HO shell the most deformed nuclear state is the one with the highest value of the $\hat C_2$ operator. Due to the the dependence of the $C_2$ on the number of the symmetric quanta $\lambda+\mu$ as in Eq. (\ref{C2}), it happens, that usually the most deformed state has the greatest number of symmetric components. In addition, since the expression (\ref{QQ}) enters the Elliott Hamiltonian with a minus sign, the state with large $QQ$ interaction (or large $C_2$ as in Eq. (\ref{QQ})) lies lower in energy.

Consequently it would be tempting to say, that the most deformed irrep, which also has the maximum number of symmetric quanta $\lambda+\mu$, satisfies the principle of minimum energy and thus represents the ground state of the nucleus. But at section \ref{symmetric} we will argue, that the most deformed irrep is not always the most symmetric, which is preferred for describing the low--lying nuclear properties \cite{PVIWigner,Talmi}. 

\section{Particle configurations within the highest weight and the most deformed irrep}\label{particle}

Each Elliott SU(3) irrep is the result of a certain particle distribution in the valence space $\ket{n_z,n_x,n_y}$. This space along with the spinor $\ket{n_z,n_x,n_y,m_s}$ transforms to the usual Shell Model orbitals as in \cite{proxy4}. It is interesting to trace back the particle configuration, which corresponds to the hw irreps and to the most deformed irreps.

Let for instance 10 protons, to be distributed in the $pf$ shell with $U(10)$ symmetry. This shell consists of the spatial orbitals, which are presented in (\ref{order3}). The highest weight irrep, which happens to be the most deformed too for this example, according to the weight vector of Eq. (\ref{w10}) is being derived, if the 10 protons occupy the states (see Eq. (\ref{order3})):
\begin{gather}\label{old}
\ket{n_z,n_x,n_y}:\ket{3,0,0},\ket{2,1,0},\ket{2,0,1},\ket{1,2,0},\ket{1,1,1}.
\end{gather}
The resulting irrep is $(10,4)$.

The addition of two more protons in this shell may result to the hw $(12,0)$ irrep. This irrep for 12 particles in $pf$ shell is being derived, if these two more protons occupy one of the empty orbitals (see Eq. (\ref{order3})):
\begin{gather}\label{empty}
\ket{n_z,n_x,n_y}:\ket{1,0,2},\ket{0,3,0},\ket{0,2,1},\ket{0,1,2},\ket{0,0,3}
\end{gather}
and specifically if the newcomers occupy the spatial orbital $\ket{1,0,2}$, while the 10 previously placed  protons remain in the orbitals of Eq. (\ref{old}), as they were. Such a behavior is in accordance with the Pauli principle in the manner, that the newcomers are subjected to a repulsive nucleon--nucleon interaction, as outlined in section \ref{force}, when they attempt to occupy an already filled spatial orbital. This is the Pauli blocking effect, which is responsible for the repulsive core at extremely short distances in all the effective NN  potentials \cite{Aoki2010}. 

The most deformed irrep $(4,10)$ however is being derived, if the 12 protons occupy the orbitals:
\begin{gather}
\ket{n_z,n_x,n_y}:\ket{1,1,1},\ket{1,0,2},\ket{0,3,0},\nonumber\\
\ket{0,2,1},\ket{0,1,2},\ket{0,0,3},
\end{gather} 
which means, that when the newcomers occupy the empty orbital $\ket{1,0,2}$ according to the Pauli blocking effect, only two protons remain in the already filled orbital $\ket{1,1,1}$, while there is an unexpected knockout of 8 protons from the orbitals $\ket{3,0,0}$, $\ket{2,1,0}$, $\ket{2,0,1}$, $\ket{1,2,0}$ to the $\ket{0,3,0}$, $\ket{0,2,1}$, $\ket{0,1,2}$, $\ket{0,0,3}$ respectively. This particle knockout could only be justified by the Principle of Minimum Energy in the sense, that this particle configuration maximizes the $QQ$ interaction, which in turn minimizes the energy. But the knockout of 8 particles by the 2 newcomers cannot be justified by any short--range NN interaction. Furthermore this most deformed irrep $(4,10)$ contains $\mu=10$ non symmetric quanta, while the highest weight irrep  $(12,0)$ of this example is totally symmetric. The most symmetric irrep, despite of the fact that it is not corresponding to the largest value of $QQ$, prevails \cite{proxy2,pseudohw}. This preference to the hw irrep stems from the short--range attractive NN interaction, which favors the maximum spatial overlapping among the nucleons \cite{PVIWigner,Talmi}. This simple example shows, that while filling the shell with particles, the highest weight irreps correspond to smooth particle distributions, without particle knockouts. On the contrary the most deformed irrep is accompanied by sudden particle displacements just after the mid--shell region.

The number of symmetric components in the spatial SU(3) wave function is $\lambda+\mu$, as explained in section \ref{sec1}. Consequently as Elliott observed in the ``Conclusions" of the introductory publication of the Elliott SU(3) symmetry \cite{Elliott1}, in the U(3) classification scheme of the $sd$ shell the highest weight irrep is the most symmetric and lies lowest in energy. This may be considered to be a general property of deformed nuclei, which stems from the attractive, short range NN interaction. Furthermore in section \ref{symmetric} we will outline, that not only the number of symmetric components $\lambda+\mu$ of the spatial SU(3) irrep is of high importance, but also the percentage $r$, as introduced in Eq. (\ref{r}), is a measure of the symmetry of the wave function.

\section{The favored SU(3) irrep}\label{symmetric}

The question is, ``which is the most spatially symmetric SU(3) irrep?", which is favored by the attractive short range interaction. For a certain number of valence protons or neutrons in the level of the $U(\Omega)$ symmetry all the possible irreps with two identical particles in each of the filled orbitals have the same eigenvalue of the $P^x$ operator of Eq. (\ref{P}), which applies for the permutation of particles. A distinction, about which of them is the most spatially symmetric irrep, can only be accomplished at the level of the SU(3) symmetry, where the permutation of quanta (not particles anymore) can be determined through the irreps $(\lambda,\mu)$.

For a certain number of valence protons/neutrons in a certain 3D-HO shell an irrep $(\lambda',\mu')$ has more symmetric components than a irrep $(\lambda,\mu)$ if:
\begin{equation}
\lambda'+\mu'>\lambda+\mu.
\end{equation}
But since $\mu$ represents the components, which are neither symmetric nor antisymmetric, the percentage $r$ of Eq. (\ref{r}) has to be also considered. Therefore we propose the two--fold condition:\\
{\it For a given number of particles in a given 3D-HO shell an SU(3) irrep $(\lambda',\mu')$ is more symmetric than an irrep $(\lambda,\mu)$}:
\begin{gather}
\mbox{if }\lambda'+\mu'\ge \lambda+\mu,\label{1}\\
\mbox{and if }r(\lambda',\mu')>r(\lambda,\mu),\label{2}
\end{gather}
where $r$ is defined in Eq. (\ref{r}). Using this two--fold condition we shall check, if there is a more symmetric Elliott SU(3) irrep $(\lambda',\mu')$ than the highest weight irrep $(\lambda,\mu)$ for every number of valence protons/neutrons in each valence 3D-HO shell possessing a $U(\Omega)$ symmetry with $\Omega=6,10,15$. The irreps in Tables \ref{U(6)}-\ref{U(15)d} are ordered in decreasing weight, thus for a given number of valence particles the highest weight irrep $(\lambda,\mu)$ is presented first. Only the irreps $(\lambda',\mu')$ with $\lambda'+\mu'\ge \lambda+\mu$ are considered according to the condition (\ref{1}).

\begin{table}
\caption{Part of the Elliott SU(3) irreps, which result from the reduction $U(6)\supset SU(3)$ \cite{code, Assimakis}. These irreps apply for the 3D-HO shell among magic numbers 8-20 \cite{Elliott1,Elliott2}, or for the proxy-SU(3) shell among magic numbers 6-12 \cite{proxy4}. For a certain number of valence particles the highest weight irrep $(\lambda,\mu)$ is presented first. The rest irreps $(\lambda',\mu')$ (for the same number of valence particles) follow with decreasing weight. Only the irreps with $\lambda'+\mu'\ge \lambda+\mu$ are presented, according to the condition (\ref{1}). In the last column the ratio $r$ as introduced in Eq. (\ref{r}), which is the percentage of the symmetric quanta out of the total in an Elliott or proxy-SU(3) wave function, is presented. It turns out, that for a certain number of valence particles, no irrep satisfies simultaneously the two conditions (\ref{1}), (\ref{2}), when comparing with the highest weight irrep. Furthermore for 7 valence particles the irrep $(\lambda',\mu')=(1,5)$ has the same number of symmetric quanta as the highest weight irrep $(\lambda,\mu)=(4,2)$  ($\lambda'+\mu'=\lambda+\mu$) and is more deformed $C_2'>C_2$, but contains less percentage of symmetric quanta $r(\lambda',\mu')<r(\lambda,\mu)$. As a result this most deformed irrep is not more symmetric than the highest weight irrep according to the hypotheses (\ref{1}) and (\ref{2}). }\label{U(6)}
\begin{tabular}{lll}
\[\begin{array}{ccccc}
 \text{ valence particles} & \lambda & \mu & C_2& \mbox{r }(\%)  \\
 1 & 2 & 0 & 10 & 100  \\
 \text{} & \text{} & \text{} & \text{} \\
 2 & 4 & 0 & 28 & 100 \\
 \text{} & \text{} & \text{} & \text{}  \\
 3 & 4 & 1 & 36 & 83  \\
 \text{} & 2 & 2 & 24 & 67  \\
 \text{} & \text{} & \text{} & \text{}  \\
 4 & 4 & 2 & 46 & 75  \\
 \text{} & \text{} & \text{} & \text{} \\
 5 & 5 & 1 & 49 & 86 \\
 \text{} & 2 & 4 & 46 & 60  \\
 \text{} & \text{} & \text{} & \text{} \\
 6 & 6 & 0 & 54 & 100\\
 \text{} & 3 & 3 & 45 & 67 \\
 \text{} & 0 & 6 & 54 & 50 \\
 \text{} & \text{} & \text{} & \text{} \\
 7 & 4 & 2 & 46 & 75  \\
 \text{} & 1 & 5 & 49 & 54 \\
 \text{} & \text{} & \text{} & \text{} \\
 8 & 2 & 4 & 46 & 60 \\
 \text{} & \text{} & \text{} & \text{}  \\
 9 & 1 & 4 & 36 & 55 \\
 \text{} & \text{} & \text{} & \text{}  \\
 10 & 0 & 4 & 28 & 50 \\
 \text{} & \text{} & \text{} & \text{} \\
 11 & 0 & 2 & 10 & 50 \\
\end{array}\]
\end{tabular}
\end{table}

\begin{table}
\caption{The same as Table \ref{U(6)} but for the $U(10)$ symmetry, which applies for the 3D-HO magic numbers 20-40 for the Elliott SU(3) symmetry, or for the 28-48 magic numbers for the proxy-SU(3) symmetry \cite{proxy4}. The irreps are presented again in decreasing weight for a given number of valence particles. Only the irreps with more symmetric quanta than those of the highest weight irrep are shown (see hypothesis (\ref{1})). }\label{U(10)a}
\begin{tabular}{lll}
\[\begin{array}{ccccc}
 \text{ valence particles} & \lambda & \mu & C_2& \mbox{r }(\%)  \\
  1 & 3 & 0 & 18 & 100 \\
 \text{} & \text{} & \text{} & \text{} \\
 2 & 6 & 0 & 54 & 100 \\
 \text{} & \text{} & \text{} & \text{} \\
 3 & 7 & 1 & 81 & 89 \\
 \text{} & \text{} & \text{} & \text{} \\
 4 & 8 & 2 & 114 & 83 \\
 \text{} & \text{} & \text{} & \text{} \\
 5 & 10 & 1 & 144 & 92 \\
 \text{} & 7 & 4 & 126 & 73 \\
 \text{} & \text{} & \text{} & \text{} \\
 6 & 12 & 0 & 180 & 100 \\
 \text{} & 9 & 3 & 153 & 80 \\
 \text{} & 6 & 6 & 144 & 67 \\
 \text{} & \text{} & \text{} & \text{} \\
 7 & 11 & 2 & 186 & 87 \\
 \text{} & 8 & 5 & 168 & 72 \\
 \text{} & \text{} & \text{} & \text{} \\
 8 & 10 & 4 & 198 & 78 \\
 \text{} & \text{} & \text{} & \text{} \\
 9 & 10 & 4 & 198 & 78 \\
 \text{} & 7 & 7 & 189 & 67 \\
 \text{} & \text{} & \text{} & \text{} \\
 10 & 10 & 4 & 198 & 78 \\
 \text{} & 7 & 7 & 189 & 67 \\
 \text{} & 4 & 10 & 198 & 58 \\
 \text{} & \text{} & \text{} & \text{} \\
\end{array}\]
\end{tabular}
\end{table}

\begin{table}
\caption{Continuation of Table \ref{U(10)a}. For 11-15 valence particles the most deformed irrep is other than the highest weight and has less percentage of symmetric quanta out of the total from the highest weight irrep ($r'<r$).}\label{U(10)b}
\begin{tabular}{lll}
\[\begin{array}{ccccc}
 \text{ valence particles} & \lambda & \mu & C_2& \mbox{r }(\%)  \\
  11 & 11 & 2 & 186 & 87 \\
 \text{} & 7 & 7 & 189 & 67 \\
 \text{} & 8 & 5 & 168 & 72 \\
 \text{} & 4 & 10 & 198 & 58 \\
 \text{} & 5 & 8 & 168 & 62 \\
 \text{} & 2 & 11 & 186 & 54 \\
 \text{} & \text{} & \text{} & \text{} \\
 12 & 12 & 0 & 180 & 100 \\
 \text{} & 8 & 5 & 168 & 72 \\
 \text{} & 9 & 3 & 153 & 80 \\
 \text{} & 4 & 10 & 198 & 58 \\
 \text{} & 5 & 8 & 168 & 62 \\
 \text{} & 6 & 6 & 144 & 67 \\
 \text{} & 3 & 9 & 153 & 57 \\
 \text{} & 0 & 12 & 180 & 50 \\
 \text{} & \text{} & \text{} & \text{} \\
 13 & 9 & 3 & 153 & 80 \\
 \text{} & 5 & 8 & 168 & 62 \\
 \text{} & 6 & 6 & 144 & 67 \\
 \text{} & 2 & 11 & 186 & 54 \\
 \text{} & 3 & 9 & 153 & 57 \\
 \text{} & \text{} & \text{} & \text{} \\
 14 & 6 & 6 & 144 & 67 \\
 \text{} & 3 & 9 & 153 & 57 \\
 \text{} & 0 & 12 & 180 & 50 \\
 \text{} & \text{} & \text{} & \text{} \\
 15 & 4 & 7 & 126 & 61 \\
 \text{} & 1 & 10 & 144 & 52 \\
 \text{} & \text{} & \text{} & \text{} \\
 16 & 2 & 8 & 114 & 56 \\
 \text{} & \text{} & \text{} & \text{} \\
 17 & 1 & 7 & 81 & 53 \\
 \text{} & \text{} & \text{} & \text{} \\
 18 & 0 & 6 & 54 & 50 \\
 \text{} & \text{} & \text{} & \text{} \\
 19 & 0 & 3 & 18 & 50 \\
\end{array}\]
\end{tabular}
\end{table}

\begin{table}
\caption{The same as Table \ref{U(6)} but for the $U(15)$, which applies for the 3D-HO magic numbers 40-70 for the Elliott SU(3) symmetry, or for the 50-80 magic numbers for the proxy-SU(3) symmetry \cite{proxy4}. }\label{U(15)a}
\begin{tabular}{lll}
\[\begin{array}{ccccc}
 \text{ valence particles} & \lambda & \mu & C_2& \mbox{r }(\%)  \\
  1 & 4 & 0 & 28 & 100 \\
 \text{} & \text{} & \text{} & \text{} & \text{} \\
 2 & 8 & 0 & 88 & 100 \\
 \text{} & \text{} & \text{} & \text{} & \text{} \\
 3 & 10 & 1 & 144 & 92 \\
 \text{} & \text{} & \text{} & \text{} & \text{} \\
 4 & 12 & 2 & 214 & 88 \\
 \text{} & \text{} & \text{} & \text{} & \text{} \\
 5 & 15 & 1 & 289 & 94 \\
 \text{} & 12 & 4 & 256 & 80 \\
 \text{} & \text{} & \text{} & \text{} & \text{} \\
 6 & 18 & 0 & 378 & 100 \\
 \text{} & 15 & 3 & 333 & 86 \\
 \text{} & \text{} & \text{} & \text{} & \text{} \\
 7 & 18 & 2 & 424 & 91 \\
 \text{} & 15 & 5 & 385 & 80 \\
 \text{} & \text{} & \text{} & \text{} & \text{} \\
 8 & 18 & 4 & 478 & 85 \\
 \text{} & \text{} & \text{} & \text{} & \text{} \\
 9 & 19 & 4 & 522 & 85 \\
 \text{} & 16 & 7 & 486 & 77 \\
 \text{} & \text{} & \text{} & \text{} & \text{} \\
 10 & 20 & 4 & 568 & 86 \\
 \text{} & 17 & 7 & 529 & 77 \\
 \text{} & 14 & 10 & 508 & 71 \\
 \text{} & \text{} & \text{} & \text{} & \text{} \\
 11 & 22 & 2 & 604 & 92 \\
 \text{} & 18 & 7 & 574 & 78 \\
 \text{} & 19 & 5 & 553 & 83 \\
 \text{} & 15 & 10 & 550 & 71 \\
 \text{} & 16 & 8 & 520 & 75 \\
 \text{} & 13 & 11 & 505 & 69 \\
 \text{} & 10 & 14 & 508 & 63 \\
 \text{} & \text{} & \text{} & \text{} & \text{} \\
 12 & 24 & 0 & 648 & 100 \\
 \text{} & 20 & 5 & 600 & 83 \\
 \text{} & 21 & 3 & 585 & 89 \\
 \text{} & 16 & 10 & 594 & 72 \\
 \text{} & 17 & 8 & 564 & 76 \\
 \text{} & 18 & 6 & 540 & 80 \\
 \text{} & 14 & 11 & 546 & 69 \\
 \text{} & 15 & 9 & 513 & 73 \\
 \text{} & 11 & 14 & 546 & 64 \\
 \text{} & 12 & 12 & 504 & 67 \\
 \text{} & 9 & 15 & 513 & 62 \\
 \text{} & 6 & 18 & 540 & 57 \\
 \text{} & \text{} & \text{} & \text{} & \text{} \\
 13 & 22 & 3 & 634 & 89 \\
 \text{} & 18 & 8 & 610 & 76 \\
 \text{} & 19 & 6 & 586 & 81 \\
 \text{} & 15 & 11 & 589 & 70 \\
 \text{} & 16 & 9 & 556 & 74 \\
 \text{} & 12 & 14 & 586 & 65 \\
 \text{} & 13 & 12 & 544 & 68 \\
 \text{} & 10 & 15 & 550 & 63 \\
 \text{} & 7 & 18 & 574 & 58 \\
 \text{} & \text{} & \text{} & \text{} & \text{} \\
 14 & 20 & 6 & 634 & 81 \\
 \text{} & 17 & 9 & 601 & 74 \\
 \text{} & 14 & 12 & 586 & 68 \\
 \text{} & 11 & 15 & 589 & 63 \\
 \text{} & 8 & 18 & 610 & 59 \\
\end{array}\]
\end{tabular}
\end{table}

\begin{table}
\caption{Continuation of Table \ref{U(15)a}. For 16-19 valence particles the most deformed irrep is other than the highest weight. No irrep $(\lambda',\mu')$, including the most deformed, satisfies simultaneously the hypotheses (\ref{1}) and (\ref{2}), when competing in symmetry with the highest weight irrep $(\lambda,\mu)$.  }\label{U(15)b}
\begin{tabular}{lll}
\[\begin{array}{ccccc}
 \text{ valence particles} & \lambda & \mu & C_2& \mbox{r }(\%)  \\
  15 & 19 & 7 & 621 & 79 \\
 \text{} & 16 & 10 & 594 & 72 \\
 \text{} & 13 & 13 & 585 & 67 \\
 \text{} & 10 & 16 & 594 & 62 \\
 \text{} & 7 & 19 & 621 & 58 \\
 \text{} & \text{} & \text{} & \text{} & \text{} \\
 16 & 18 & 8 & 610 & 76 \\
 \text{} & 15 & 11 & 589 & 70 \\
 \text{} & 12 & 14 & 586 & 65 \\
 \text{} & 9 & 17 & 601 & 60 \\
 \text{} & 6 & 20 & 634 & 57 \\
 \text{} & \text{} & \text{} & \text{} & \text{} \\
 17 & 18 & 7 & 574 & 78 \\
 \text{} & 14 & 12 & 586 & 68 \\
 \text{} & 15 & 10 & 550 & 71 \\
 \text{} & 11 & 15 & 589 & 63 \\
 \text{} & 12 & 13 & 544 & 66 \\
 \text{} & 8 & 18 & 610 & 59 \\
 \text{} & 9 & 16 & 556 & 61 \\
 \text{} & 6 & 19 & 586 & 57 \\
 \text{} & 3 & 22 & 634 & 53 \\
 \text{} & \text{} & \text{} & \text{} & \text{} \\
 18 & 18 & 6 & 540 & 80 \\
 \text{} & 14 & 11 & 546 & 69 \\
 \text{} & 15 & 9 & 513 & 73 \\
 \text{} & 10 & 16 & 594 & 62 \\
 \text{} & 11 & 14 & 546 & 64 \\
 \text{} & 12 & 12 & 504 & 67 \\
 \text{} & 8 & 17 & 564 & 60 \\
 \text{} & 9 & 15 & 513 & 62 \\
 \text{} & 5 & 20 & 600 & 56 \\
 \text{} & 6 & 18 & 540 & 57 \\
 \text{} & 3 & 21 & 585 & 53 \\
 \text{} & 0 & 24 & 648 & 50 \\
 \text{} & \text{} & \text{} & \text{} & \text{} \\
 19 & 19 & 3 & 493 & 88 \\
 \text{} & 14 & 10 & 508 & 71 \\
 \text{} & 15 & 8 & 478 & 74 \\
 \text{} & 16 & 6 & 454 & 79 \\
 \text{} & 10 & 15 & 550 & 63 \\
 \text{} & 11 & 13 & 505 & 65 \\
 \text{} & 12 & 11 & 466 & 68 \\
 \text{} & 13 & 9 & 433 & 71 \\
 \text{} & 7 & 18 & 574 & 58 \\
 \text{} & 8 & 16 & 520 & 60 \\
 \text{} & 9 & 14 & 472 & 62 \\
 \text{} & 10 & 12 & 430 & 65 \\
 \text{} & 5 & 19 & 553 & 56 \\
 \text{} & 6 & 17 & 496 & 58 \\
 \text{} & 7 & 15 & 445 & 59 \\
 \text{} & 2 & 22 & 604 & 52 \\
 \text{} & 3 & 20 & 538 & 53 \\
 \text{} & 4 & 18 & 478 & 55 \\
 \text{} & 1 & 21 & 529 & 51 \\
\end{array}\]
\end{tabular}
\end{table}

\begin{table}
\caption{Continuation of Tables \ref{U(15)a} and \ref{U(15)b}. For 20-23 valence particles no irrep is more symmetric than the highest weight according to the hypotheses (\ref{1}) and (\ref{2}). }\label{U(15)c}
\begin{tabular}{lll}
\[\begin{array}{ccccc}
 \text{ valence particles} & \lambda & \mu & C_2& \mbox{r }(\%)  \\
   20 & 20 & 0 & 460 & 100 \\
 \text{} & 15 & 7 & 445 & 76 \\
 \text{} & 16 & 5 & 424 & 81 \\
 \text{} & 17 & 3 & 409 & 87 \\
 \text{} & 10 & 14 & 508 & 63 \\
 \text{} & 11 & 12 & 466 & 66 \\
 \text{} & 12 & 10 & 430 & 69 \\
 \text{} & 13 & 8 & 400 & 72 \\
 \text{} & 14 & 6 & 376 & 77 \\
 \text{} & 7 & 17 & 529 & 59 \\
 \text{} & 8 & 15 & 478 & 61 \\
 \text{} & 9 & 13 & 433 & 63 \\
 \text{} & 10 & 11 & 394 & 66 \\
 \text{} & 11 & 9 & 361 & 69 \\
 \text{} & 4 & 20 & 568 & 55 \\
 \text{} & 5 & 18 & 508 & 56 \\
 \text{} & 6 & 16 & 454 & 58 \\
 \text{} & 7 & 14 & 406 & 60 \\
 \text{} & 8 & 12 & 364 & 63 \\
 \text{} & 3 & 19 & 493 & 54 \\
 \text{} & 4 & 17 & 436 & 55 \\
 \text{} & 5 & 15 & 385 & 57 \\
 \text{} & 0 & 22 & 550 & 50 \\
 \text{} & 1 & 20 & 484 & 51 \\
 \text{} & 2 & 18 & 424 & 53 \\
 \text{} & \text{} & \text{} & \text{} & \text{} \\
 21 & 16 & 4 & 396 & 83 \\
 \text{} & 11 & 11 & 429 & 67 \\
 \text{} & 12 & 9 & 396 & 70 \\
 \text{} & 13 & 7 & 369 & 74 \\
 \text{} & 7 & 16 & 486 & 59 \\
 \text{} & 8 & 14 & 438 & 61 \\
 \text{} & 9 & 12 & 396 & 64 \\
 \text{} & 10 & 10 & 360 & 67 \\
 \text{} & 4 & 19 & 522 & 55 \\
 \text{} & 5 & 17 & 465 & 56 \\
 \text{} & 6 & 15 & 414 & 58 \\
 \text{} & 7 & 13 & 369 & 61 \\
 \text{} & 2 & 20 & 510 & 52 \\
 \text{} & 3 & 18 & 450 & 54 \\
 \text{} & 4 & 16 & 396 & 56 \\
 \text{} & 1 & 19 & 441 & 51 \\
 \text{} & \text{} & \text{} & \text{} & \text{} \\
 22 & 12 & 8 & 364 & 71 \\
 \text{} & 8 & 13 & 400 & 62 \\
 \text{} & 9 & 11 & 361 & 65 \\
 \text{} & 4 & 18 & 478 & 55 \\
 \text{} & 5 & 16 & 424 & 57 \\
 \text{} & 6 & 14 & 376 & 59 \\
 \text{} & 3 & 17 & 409 & 54 \\
 \text{} & 0 & 20 & 460 & 50 \\
 \text{} & \text{} & \text{} & \text{} & \text{} \\
 23 & 9 & 10 & 328 & 66 \\
 \text{} & 5 & 15 & 385 & 57 \\
 \text{} & 6 & 13 & 340 & 59 \\
 \text{} & 2 & 18 & 424 & 53 \\
 \text{} & 3 & 16 & 370 & 54 \\
 \text{} & \text{} & \text{} & \text{} & \text{} \\
\end{array}\]
\end{tabular}
\end{table}

\begin{table}
\caption{Continuation of Tables \ref{U(15)a}, \ref{U(15)b} and \ref{U(15)c}. No irrep competes in symmetry the highest weight irrep according to conditions (\ref{1}) and (\ref{2}). }\label{U(15)d}
\begin{tabular}{lll}
\[\begin{array}{ccccc}
 \text{ valence particles} & \lambda & \mu & C_2& \mbox{r }(\%)  \\
 24 & 6 & 12 & 306 & 60 \\
 \text{} & 3 & 15 & 333 & 55 \\
 \text{} & 0 & 18 & 378 & 50 \\
 \text{} & \text{} & \text{} & \text{} & \text{} \\
 25 & 4 & 12 & 256 & 57 \\
 \text{} & 1 & 15 & 289 & 52 \\
 \text{} & \text{} & \text{} & \text{} & \text{} \\
 26 & 2 & 12 & 214 & 54 \\
 \text{} & \text{} & \text{} & \text{} & \text{} \\
 27 & 1 & 10 & 144 & 52 \\
 \text{} & \text{} & \text{} & \text{} & \text{} \\
 28 & 0 & 8 & 88 & 50 \\
 \text{} & \text{} & \text{} & \text{} & \text{} \\
 29 & 0 & 4 & 28 & 50 \\
\end{array}\]
\end{tabular}
\end{table}

From Tables \ref{U(6)}-\ref{U(15)d} it emerges, that no irrep satisfies simultaneously the two conditions (\ref{1}), (\ref{2}), when competing with the highest weight irrep. The irreps of the $U(21)$ symmetry, which applies for the 3D-HO shell among magic numbers 70-112 in the Elliott SU(3) symmetry or for the 82-124 shell in the proxy-SU(3) symmetry \cite{proxy4}, have also been checked, but have not been presented here, because they are too lengthy. The same conclusion applies for all the shells with $U(6)$, $U(10)$, $U(15)$, $U(21)$ symmetry: according to the conditions (\ref{1}) and (\ref{2}) there is no irrep $(\lambda',\mu')$, which is more symmetric, than the highest weight irrep $(\lambda,\mu)$. From {\it reductio ad absurdum}, we may state, that for any number of valence particles in any valence shell, the highest weight irrep is the most symmetric among the rest possible ones. It seems, that this irrep has the finest balance among the maximization of the symmetric quanta $\lambda+\mu$ along with the minimization of the non symmetric quanta $\mu$.

As the authors of Ref. \cite{PVIWigner} have enunciated, the most favorable spatial SU(3) irrep, is the most symmetric among all the possible ones. This conclusion stems right from the short--range character of the attractive NN interaction, which favors the maximal spatial overlap among the fermions. Consequently the hw irrep is the {\it favored} one and describes best the low--lying nuclear properties \cite{proxy2,pseudohw}.

\section{The particle-hole symmetry}\label{particle-hole}

The highest weight irreps (hw) and those, which correspond to the maximum value of the $C_2$ operator (C) are presented in Table I of Ref. \cite{proxy2}. If there is a particle--hole symmetry in the SU(3) irreps, then for the same number of valence particles $m_p$ and valence holes $m_h$, the relevant SU(3) irrep is produced by an interchange of $\lambda\leftrightarrow \mu$. This particle--hole symmetry for a proton or neutron 3D-HO shell leads to an interchance of $\lambda$ and $\mu$ for the particle and hole SU(3) irreps \cite{Elliott4}:
\begin{equation}\label{ph}
(\lambda,\mu)_{m_p}\rightarrow (\mu,\lambda)_{m_h=m_p}.
\end{equation}

For instance the highest weight irrep for two valence protons in the $U(6)$ is $(4,0)$, while the hw irrep for two valence holes (or 10 valence protons, since this shell may accommodate at most 12 protons) is $(0,4)$. Similar expressions may be reproduced for any other shell, if there is such a type of particle--hole symmetry. If the favored irrep was the one with the maximum value of the $C_2$ operator, then such a particle-hole symmetry would exist in the SU(3) irreps for all the shells, namely the $pf$ with $U(10)$ symmetry, the $sdg$ with $U(15)$ symmetry, etc (see Table I of Ref. \cite{proxy2}).

On the contrary the highest weight SU(3) irreps, which are presented in Table \ref{hwirreps}, sometimes do not respect this interchange of $\lambda\leftrightarrow\mu$ for the same number of particles and holes. For instance the highest weight irrep for 7 valence protons in the $sd$ shell with $U(6)$ symmetry is $(4,2)$, while the hw for 7 valence holes  (or 5 valence protons) in the same shell is $(5,1)$. Obviously the interchange of Eq. (\ref{ph}) is not always functional for the highest weight irreps. For the rest of the shells ($pf$, $sdg$, etc) this type of particle--hole asymmetry in the highest weight irreps, which are presented in Table \ref{hwirreps}, is more intense. This phenomenon has been discussed in Ref. \cite{proxy2}. In Table \ref{hwirreps} we see that most of the hw irreps listed are prolate ($\lambda>\mu$), while an oblate ($\lambda<\mu$) region appears above mid-shell. In U(6), U(10), U(15), and U(21) in particular, which can accommodate respectively up to 12, 20, 30, 42 identical nucleons, the oblate region starts at 8, 15, 23, 34 nucleons respectively.  

Nevertheless the particle--hole symmetry exists, even for the highest weight irreps, but in another way. The calculation of the hw irreps for a certain number of valence particles has been presented in section \ref{sec1}. One may calculate the hw irreps for a certain number of valence holes, by filling the spatial orbitals of Eq. (\ref{order}) in the inverse order. As an example the filling of the $sd$ shell with holes is equivalent to the filling of the $\ket{n_z,n_x,n_y}$ spatial orbitals, with the following order:
\begin{gather}
\ket{n_z,n_x,n_y}:\ket{0,0,2},\ket{0,1,1},\ket{0,2,0},\nonumber\\
\ket{1,0,1},\ket{1,1,0},\ket{2,0,0}.
\end{gather}
For instance the 7 valence holes in this shell occupy the orbitals:
 \begin{gather}
\ket{n_z,n_x,n_y}:\ket{0,0,2},\ket{0,1,1},\ket{0,2,0},\nonumber\\
\ket{1,0,1}
\end{gather}
and lead to summations of quanta as in Eq. (\ref{summations}):
\begin{equation}
\sum_{i=1}^7n_{z,i}=1, \quad \sum_{i=1}^7n_{x,i}=6, \quad \sum_{i=1}^7n_{y,i}=7,
\end{equation}
thus the relevant $U(3)$ irrep according to Eq. (\ref{fi}) is:
\begin{equation}
[f_1,f_2,f_3]=[7,6,1],
\end{equation}
which leads to the highest weight SU(3) irrep $(1,5)$ (see Eqs. (\ref{l}) and (\ref{m})). This irrep using an interchange of $\lambda\leftrightarrow\mu$, becomes $(5,1)$ for 5 valence particles in the $U(6)$, as it should be. The proton or neutron capacity of a shell, which consists of orbitals with $\mathcal{N}$ number of quanta, is $(\mathcal{N}+1)(\mathcal{N}+2)$. If the number of valence particles $m_p$ and the number of valence holes $m_h$ is complementary:
\begin{equation}
m_p+m_h=(\mathcal{N}+1)(\mathcal{N}+2),
\end{equation}
then the particle and hole hw SU(3) irreps are related by an interchange of $\lambda\leftrightarrow \mu$:
\begin{equation}
(\lambda,\mu)_{m_p}\rightarrow (\mu,\lambda)_{m_h=(\mathcal{N}+1)(\mathcal{N}+2)-m_p}.
\end{equation}

\begin{table}
\centering
\caption{Highest weight SU(3) irreps (which always have multiplicity one) for $U(\Omega)$, $\Omega=6, 10, 15, 21$ for $m_p$ valence protons or neutrons, derived using the code UNTOU3 \cite{code}. Violations of the particle--hole symmetry, as expressed in Eq. (\ref{ph}), appearing in the lower half of each column are indicated by boldface characters.}\label{hwirreps} 
\begin{tabular}{ r  r r r r   }
\hline\noalign{\smallskip}
$m_p$  & $U(6)$      & $U(10)$     & $U(15)$      & $U(21)$      \\
\noalign{\smallskip}\hline\noalign{\smallskip}
 0 &(0,0)      &(0,0)      &(0,0)       &(0,0)       \\  
 1 &(2,0)      &(3,0)      & (4,0)      & (5,0)      \\
 2 &(4,0)      & (6,0)     & (8,0)      &(10,0)      \\
 3 &(4,1)      & (7,1)     &(10,1)      &(13,1)      \\
 4 &(4,2)      & (8,2)     &(12,2)      &(16,2)      \\
 5 &(5,1)      &(10,1)     &(15,1)      &(20,1)      \\
 6 &(6,0)      &(12,0)     &(18,0)      &(24,0)      \\
 7 &{\bf(4,2)} &(11,2)     &(18,2)      &(25,2)      \\
 8 &(2,4)      &(10,4)     &(18,4)      &(26,4)      \\
 9 &(1,4)      &(10,4)     &(19,4)      &(28,4)      \\
10 &(0,4)      &(10,4)     &(20,4)      &(30,4)      \\
11 &(0,2)      &{\bf(11,2)}&(22,2)      &(33,2)      \\
12 &(0,0)      &{\bf(12,0)}&(24,0)      &(36,0)      \\
13 &           &{\bf(9,3)} &(22,3)      &(35,3)      \\
14 &           &{\bf(6,6)} &(20,6)      &(34,6)      \\
15 &           &{\bf(4,7)} &(19,7)      &(34,7)      \\
16 &           & (2,8)     &{\bf(18,8)} &(34,8)      \\
17 &           & (1,7)     &{\bf(18,7)} &(35,7)      \\
18 &           & (0,6)     &{\bf(18,6)} &(36,6)      \\
19 &           & (0,3)     &{\bf(19,3)} &(38,3)      \\
20 &           & (0,0)     &{\bf(20,0)} &(40,0)      \\
21 &           &           &{\bf(16,4)} &(37,4)      \\
22 &           &           &{\bf(12,8)} &{\bf(34,8)} \\
23 &           &           &{\bf(9,10)} &{\bf(32,10)}\\
24 &           &           &{\bf(6,12)} &{\bf(30,12)}\\
25 &           &           &{\bf(4,12)} &{\bf(29,12)}\\
26 &           &           &(2,12)      &{\bf(28,12)}\\
27 &           &           &(1,10)      &{\bf(28,10)}\\
28 &           &           & (0,8)      &{\bf(28,8)} \\
29 &           &           & (0,4)      &{\bf(29,4)} \\
30 &           &           & (0,0)      &{\bf(30,0)} \\
31 &           &           &            &{\bf(25,5)} \\
32 &           &           &            &{\bf(20,10)}\\
33 &           &           &            &{\bf(16,13)}\\
34 &           &           &            &{\bf(12,16)}\\
35 &           &           &            &{\bf(9,17)} \\ 
36 &           &           &            &{\bf(6,18)} \\
37 &           &           &            &{\bf(4,17)} \\
38 &           &           &            &(2,16)      \\
39 &           &           &            &(1,13)      \\
40 &           &           &            &(0,10)      \\
41 &           &           &            &(0,5)       \\
42 &           &           &            &(0,0)       \\
\noalign{\smallskip}\hline
\end{tabular}
\end{table}

\section{The prolate dominance in atomic nuclei}\label{prolate}

The consequences in atomic nuclei of the appearance of a majority of prolate irreps in Table \ref{hwirreps} have been studied in the framework of the proxy-SU(3) model \cite{proxy1,proxy2,proxy3}, in which the SU(3) symmetry of the harmonic oscillator shells \cite{Wybourne,Moshinsky} is extended beyond the $sd$ nuclear shell by an approximation \cite{proxy1,proxy4} involving the intruder orbitals of opposite parity within each shell. The hw irreps corresponding to the valence protons and to the valence neutrons are combined in order to provide the SU(3) irrep characterizing the whole nucleus \cite{proxy2}. 

It turns out that a prolate to oblate shape transition is predicted with the use of the hw irreps \cite{proxy2} when both protons and neutrons are near the end of the corresponding shell, thus represented by oblate ($\lambda<\mu$) SU(3) irreps. Agreement with existing experimental information \cite{Namenson,Alkhomashi,Wheldon,Podolyak,Linnemann} in the heavy rare earths, below 82 protons and 126 neutrons, has been seen. In other words we see in the nuclear chart, below the doubly magic nucleus $^{208}_{82}$Pb$_{126}$, a relatively small region of oblate nuclei, while prolate shapes are obtained everywhere else in the rare earths with 50-82 protons and 126-184 neutrons. 

A similar picture is predicted \cite{proxy2} in other regions of  the nuclear chart, for example the rare earths with 50-82 protons and 50-82 neutrons. As a consequence, the prolate over oblate dominance in the shapes of the ground state bands of even--even nuclei, which has been an open problem for many years \cite{Hamamoto}, is obtained as a direct consequence of the proxy-SU(3) symmetry and the use of the highest weight irreps \cite{proxy2,proxy3}.    

Recent studies \cite{BonRila19,pseudohw} indicate, that the prolate to oblate shape transition and the prolate over oblate dominance in the shapes of the ground state bands of even nuclei can be also obtained within the framework of the pseudo-SU(3) model \cite{pseudo1,pseudo2,DW1}, in which the normal parity orbitals in a given nuclear shell are modified through a unitary transformation \cite{Quesne}, in contrast to the proxy-SU(3) model \cite{proxy4}, in which a unitary transformation is applied to the intruder parity orbitals. In both cases the aim of the unitary transformation is the restoration of the SU(3) symmetry of the 3D-HO \cite{Wybourne,Moshinsky}, which is broken by the spin--orbit interaction beyond the $sd$ nuclear shell \cite{Mayer1}. The compatibility of the pseudo-SU(3) and proxy-SU(3) approximations has also been demonstrated recently in the study of quarteting in heavy nuclei \cite{Cseh}.  

It should be emphasized, that the above findings in atomic nuclei are rooted in the attractive, short range nature of the NN interaction \cite{Ring,Casten}, which favors maximal spatial overlaps \cite{PVIWigner}. These are obtained when the spatial part of the wave function is as symmetric as possible \cite{Wigner1937,PVIWigner}. In contrast, the spin--isospin part of the wave function possesses a Young diagram which is the conjugate of the Young diagram of the spatial part, as pointed out in section \ref{wf}, so that the fully antisymmetric character of the total wave function is guaranteed. 

\begin{table*}
\centering
\caption{Experimental magic numbers for Na clusters by Martin {\it et al.} \cite{Martin1,Martin2} (column 1), 
Bj{\o}rnholm {\it et al.} \cite{Bjorn1,Bjorn2} (column 2), 
Knight {\it et al.} \cite{Knight1} (column 3), and 
Pedersen {\it et al.} \cite{Peder} (column 4), as well as to the experimental data 
for Li clusters by Br\'echignac {\it et al.} (\cite{Brec1} in column 5, 
\cite{Brec2} in column 6) are compared to theoretical predictions \cite{Mayer} by the (non--deformed) 3D-HO (column 9), the square well potential (SW) (column 8), a rounded
square well potential intermediate between the previous two (INT) (column 7), and the 3D $q$--deformed harmonic 
oscillator (DHO) \cite{Karoussos,Lenis} (column 10). }\label{Table10}    
\begin{tabular}{c c c c c c c c c c}
\hline\noalign{\smallskip}
  exp.  & exp. &  exp.&  exp.& exp. & exp.  & th.  &  th. &  th. & th.  \\
 Na  &  Na & Na &  Na  &  Li & Li & INT & SW & HO  & DHO  \\
\cite{Martin1,Martin2} &\cite{Bjorn1,Bjorn2} &\cite{Knight1} &\cite{Peder}&\cite{Brec1} &\cite{Brec2} &\cite{Mayer} &\cite{Mayer} 
& \cite{Mayer} & \cite{Karoussos,Lenis} \\  
\noalign{\smallskip}\hline\noalign{\smallskip}
    2      &   2     & 2&      &      &   2&     2   &   2     &   2 & 2   \\
    8      &   8     & 8&      &      &   8&     8   &   8     &   8 & 8   \\
   18      &         &  &      &      &    &    18   &  18     &     & (18)\\
   20      &  20     &20&      &      &  20&    20   &  20     &  20 & 20  \\
   34      &         &  &      &      &    &    34   &  34     &     & 34  \\
   40      &  40     &40&   40 &      &  40&    40   &  40     &  40 & 40  \\
   58      &  58     &58&   58 &      &  58&    58   &  58     &     & 58  \\
           &         &  &      &      &    &  68,70  &  68     &  70 &     \\
  90,92    &  92     &92&   92 &   93 &  92&    92   & 90,92   &     & 92  \\
           &         &  &      &      &    &  106,112&  106    & 112 &     \\
  138      & 138     &  &  138 &  134 & 138&    138  &  132,138&     & 138 \\
  198$\pm$2 & 196     & &  198 &  191 & 198&    156  &   156   & 168 & 198 \\
\noalign{\smallskip}\hline
\end{tabular}

\end{table*}

\section{Manifestation of prolate to oblate transition in metal clusters}\label{SecIVb}

Structural similarities between atomic clusters \cite{deHeer,Brack,Nester,deHeer2} and atomic nuclei have been pointed out \cite{Nester,Greiner} since the early days of experimental study \cite{deHeer,deHeer2} of atomic clusters. Alkali metal clusters, in particular, exhibit magic numbers, which for few particles are similar to the 3D-HO magic numbers, while they diverge at higher particle numbers \cite{Martin1,Martin2,Bjorn1,Bjorn2,Knight1,Peder,Brec1,Brec2}. The valence electrons in alkali metal clusters are supposed to be free, thus forming shells. The major magic numbers observed in alkali metal clusters are 2, 8, 20, 40, 58, 92, \dots. Experimental data \cite{Martin1,Martin2,Peder,Brec2} and theoretical predictions \cite{Karoussos,Lenis} for magic numbers in alkali metal clusters exist up to 1500 atoms. Some data and theoretical predictions up to 200 atoms are summarized in Table \ref{Table10}. 

Prolate and oblate shapes in alkali metal clusters have been seen experimentally through optical response measurements, by looking at the intensity of the various energy peaks \cite{Borggreen,Pedersen1,Pedersen2,Haberland,Schmidt}. In this way oblate shapes have been observed below cluster sizes 20 and 40 \cite{Borggreen,Pedersen1,Pedersen2,Haberland}, while prolate shapes have been observed above cluster sizes 8 and 20 \cite{Borggreen,Pedersen1,Pedersen2,Haberland} and later on above 40 \cite{Schmidt}. In other words, a pattern with oblate shapes below magic numbers and prolate shapes above magic numbers appears at light alkali metal clusters. 

From the theoretical viewpoint, alkali metal clusters have been described \cite{Clemenger} within the Nilsson model \cite{Nilsson1,Nilsson2}, initially introduced for the description of deformed nuclei. No spin--orbit term exists in the case of alkali metal clusters, while the $l^2$ term flattening the bottom of the 3D-HO potential and making its edges sharper is still in use \cite{Clemenger}. In other words, deformation in alkali metal clusters can be described by the same model used for describing deformed nuclei. 

Deformed nuclei are also known to be described by the SU(3) symmetry, in the framework of algebraic models using bosons, as the Interacting Boson Model \cite{IA,IVI,FVI} and the Vector Boson Model \cite{Raychev,Afanasev,RR27},  or fermions, like the Symplectic Model \cite{Rosensteel,RW}, the Fermion Dynamical Symmetry Model (FDSM) \cite{FDSM}, the pseudo-SU(3) model \cite{pseudo1,pseudo2,DW1}, the quasi-SU(3) model \cite{Zuker1,Zuker2}, and, recently, the proxy-SU(3) model \cite{proxy1,proxy2,proxy3,proxy4}. Therefore it is natural to see, if certain properties for alkali metal clusters can be predicted by algebraic models used for deformed nuclei, taking into account that no spin--orbit force is present in the case of atomic clusters \cite{Clemenger}, the pairing force being also absent in this case \cite{Greiner}. 

Magic numbers for alkali metal clusters have been predicted \cite{deHeer2,Mayer} by the (non--deformed) 3D-HO, the square well potential, as well as a rounded square well potential between the previous two. Predictions are in reasonable agreement to experimental findings up to cluster size around 150, as seen in Table \ref{Table10}. Experimental magic numbers up to cluster size 1500  \cite{Martin1,Martin2,Peder,Brec2} have been reproduced by a deformed 3D-HO \cite{Karoussos,Lenis}. It is seen in Table \ref{3DHO} that the $(\mathcal{N},l)$ orbitals, characterized by the number of oscillator quanta $\mathcal{N}$ and the angular momentum $l$, preserve in the deformed 3D-HO the same order as in the non--deformed harmonic oscillator up to cluster size 70, while beyond this point mixing of orbitals with different $\mathcal{N}$ starts, since from each $\mathcal{N}$ shell the orbital with the highest angular momentum $l=\mathcal{N}$ is pushed to lower energies, thus entering shells with lower values of $\mathcal{N}$.  

In view of the above, the appearance of prolate shapes above cluster sizes 8, 20, and 40, and of oblate shapes below cluster sizes 20 and 40 is easily explained by Table \ref{hwirreps}, since the 8-20 shell corresponds to the harmonic oscillator $sd$ shell with $U(6)$ symmetry, while the 20-40 shell corresponds to the $pf$ shell with $U(10)$ symmetry. According to Table \ref{hwirreps}, prolate shapes are seen at the beginning of the shells and further up within them, i.e. starting at cluster sizes 8 and 20, while oblate shapes are seen near the end of the shells, {\it i.e.}, below cluster sizes 20 and 40. Above 40 the $sdg$ shell with U(15) symmetry is starting, thus prolate shapes are again expected, as seen experimentally \cite{Schmidt}. But this pattern of succession of prolate and oblate shapes breaks down around cluster size 70, since the sequence of 3D-HO shells is disturbed, as seen in Table \ref{3DHO}. In particular, the $(\mathcal{N},l)=(4,4)$ orbital is lowered, thus providing a magic number at 58, while the remaining orbitals (4,2) and (4,0) from the $sdg$ shell are joined by the (5,5) orbital of the $pfh$ shell, forming the magic number 92.  

We conclude that the algebraic results reported in Table \ref{hwirreps} can explain both the succession of prolate and oblate shapes seen in light alkali metal clusters respectively above and below the magic numbers 8, 20, and 40, as well as the disappearance of this pattern at higher cluster sizes, a problem which has stayed open for years \cite{Greiner}.     

It should be emphasized that the similarities between several properties of atomic nuclei and atomic clusters \cite{Nester,Greiner} is due to the similar form of the relevant potentials, which at the most elementary level are in both cases modified harmonic oscillators with flattened bottoms, namely the Nilsson model \cite{Nilsson1,Nilsson2} in atomic nuclei and the Clemenger model \cite{Clemenger} in atomic clusters, while the basic differences between the two systems are rooted in the absence of the spin-orbit and pairing interactions in the case of atomic clusters \cite{Greiner}.

\begin{table}
\centering
\caption{Energy levels  of the 3-dimensional $q$-deformed 
harmonic oscillator \cite{Karoussos,Lenis}. Each level is characterized by $\mathcal{N}$ 
(the number of vibrational quanta) and  $l$ (the angular momentum), while 
$2(2l+1)$ represents the number of particles each level can accommodate, and
 under ``total'' the total number of particles up to and including 
this level is given. Magic numbers, corresponding to large energy gaps, are reported in boldface.}
\label{3DHO}      
\begin{tabular}{r r r r }
\hline\noalign{\smallskip}
$\mathcal{N}$ & $l$ & $2(2l+1)$ & total \\
\noalign{\smallskip}\hline\noalign{\smallskip}
 0&  0&  2  &  {\bf 2}   \\
 1&  1&  6  &  {\bf 8}   \\
 2&  2& 10  & 18      \\
 2&  0&  2  & {\bf 20}   \\
 3&  3& 14  & {\bf 34}  \\
 3&  1&  6  & {\bf 40}   \\
 4&  4& 18  & {\bf 58}   \\
 4&  2& 10  & 68        \\
 4&  0&  2  & 70        \\
 5&  5& 22  & {\bf 92}  \\
 5&  3& 14  &106         \\
 6&  6& 26  &132        \\
 5&  1&  6  &{\bf 138}  \\
 6&  4& 18  &156       \\
 7&  7& 30  &186        \\
 6&  2& 10  &196       \\
 6&  0&  2  &{\bf 198}  \\
\noalign{\smallskip}\hline
\end{tabular}
\end{table}

\section{Discussion}

It was the purpose of the present work to answer the stormy question why in the proxy-SU(3) model  the highest weight irreducible representation (irrep) of SU(3) is  used instead of the irrep with the highest value of the second order Casimir operator of SU(3), which corresponds to the maximum value of the quadrupole--quadrupole interaction. The basic points of the answer are the following.

a)The attractive, short range nature of the nucleon--nucleon interaction favors wave functions with as symmetric as possible spatial part, which guarantees maximal spatial overlaps among them.  

b)It is proved that the highest weight SU(3) irrep for a given number of nucleons (protons or neutrons) in a given 3-dimensional isotropic harmonic oscillator shell possessing an SU(3) subalgebra is the irrep possessing the highest percentage of symmetrized boxes in the relevant Young diagram, i.e., it represents the most symmetric spatial state for the given system. 

The dominance of the highest weight spatial irreps has the following consequences.

1) It explains the dominance of prolate over oblate shapes in the ground states of even--even nuclei.

2) In both even--even nuclei and in alkali metal clusters it predicts a shape transition from prolate to oblate shapes beyond the mid--shell and below its closure.

3) In atomic nuclei the prolate to oblate shape transition is seen experimentally in the heavy rare earths below the doubly magic nucleus $^{208}_{82}$Pb$_{126}$.  Similar transitions are predicted in other regions of the nuclear chart, yet inaccessible by experiment.

4) In alkali metal clusters, 2) explains the existence of prolate deformations above magic numbers and oblate deformations below magic numbers up to 60 atoms, as well as the disappearance of this pattern in heavier clusters.  

It should be emphasized that the conclusions of the present study are valid for any finite fermionic system governed by attractive, short range interactions and possessing the SU(3) symmetry.


\end{document}